\documentclass[screen, sigconf]{acmart}

\AtBeginDocument{%
  }

\def\namedlabel#1#2{\begingroup
    #2%
    \def\@currentlabel{#2}%
    \phantomsection\label{#1}\endgroup
}

\setcopyright{acmlicensed}
\usepackage[normalem]{ulem}
\usepackage{amsmath}
\usepackage{booktabs}
\usepackage{tabularx}
\usepackage{graphicx}

\DeclareMathOperator{\logit}{logit}

\copyrightyear{2018}
\acmYear{2018}
\acmDOI{XXXXXXX.XXXXXXX}

\acmConference[Conference acronym 'XX]{Make sure to enter the correct
  conference title from your rights confirmation emai}{June 03--05,
  2018}{Woodstock, NY}

\acmISBN{978-1-4503-XXXX-X/18/06}

\usepackage{xcolor}
\usepackage{subcaption}

\begin{document}

\title[Balancing Domestic and Global Perspectives]{Balancing Domestic and Global Perspectives: Evaluating Dual-Calibration and LLM-Generated Nudges for Diverse News Recommendation}

\author{Ruixuan Sun}
\authornote{Both authors contributed equally to this research and flipped a coin for authorship order.}
\orcid{0000-0003-4653-0384}
\affiliation{%
 \institution{Grouplens Research, University of Minnesota}
 \city{Minneapolis}
 \state{Minnesota}
 \country{United States}}

\author{Matthew Zent}
\authornotemark[1]
\orcid{0000-0003-4555-8764}
\affiliation{%
 \institution{Grouplens Research, University of Minnesota}
 \city{Minneapolis}
 \state{Minnesota}
 \country{United States}}

\author{Minzhu Zhao}
\affiliation{%
 \institution{Grouplens Research, University of Minnesota}
 \city{Minneapolis}
 \state{Minnesota}
 \country{United States}}

\author{Thanmayee Boyapati}
\affiliation{%
 \institution{Grouplens Research, University of Minnesota}
 \city{Minneapolis}
 \state{Minnesota}
 \country{United States}}

\author{Xinyi Li}
\affiliation{%
 \institution{Northwestern University}
 \city{Evanston}
 \state{Illinois}
 \country{United States}}

\author{Joseph A. Konstan}
\affiliation{%
 \institution{Grouplens Research, University of Minnesota}
 \city{Minneapolis}
 \state{Minnesota}
 \country{United States}}

\renewcommand{\shortauthors}{Trovato et al.}
\newcommand{\lxy}[1]{\textcolor{blue}{#1}} 
\newcommand{\mjz}[1]{\textcolor{magenta}{#1}}
\definecolor{darkred}{HTML}{8B0000}  
\definecolor{darkblue}{HTML}{00008B} 

\begin{abstract}
  In this study, we applied the ``personalized diversity nudge framework'' with the goal of expanding user reading coverage in terms of news locality (i.e., domestic and world news). We designed a novel topic-locality dual calibration algorithmic nudge and a large language model-based news personalization presentation nudge, then launched a 5-week real-user study with 120 U.S. news readers on the news recommendation experiment platform POPROX. With user interaction logs and survey responses, we found that algorithmic nudges can successfully increase exposure and consumption diversity, while the impact of LLM-based presentation nudges varied. User-level topic interest is a strong predictor of user clicks, while highlighting the relevance of news articles to prior read articles outperforms generic topic-based and no personalization. We also demonstrate that longitudinal exposure to calibrated news may shift readers' reading habits to value a balanced news digest from both domestic and world articles. Our results provide direction for future work on nudging for diverse consumption in news recommendation systems.  

\end{abstract}

\ccsdesc[500]{Human-centered computing~Human computer interaction (HCI)}
\ccsdesc[500]{Information systems~Users and interactive retrieval}
\ccsdesc[500]{Information systems~Recommender systems}
\ccsdesc[500]{Computing methodologies~Natural language generation}

\keywords{News Recommendation, Recommendation Calibration, Content Personalization}

\begin{teaserfigure}
    \includegraphics[width=\textwidth]{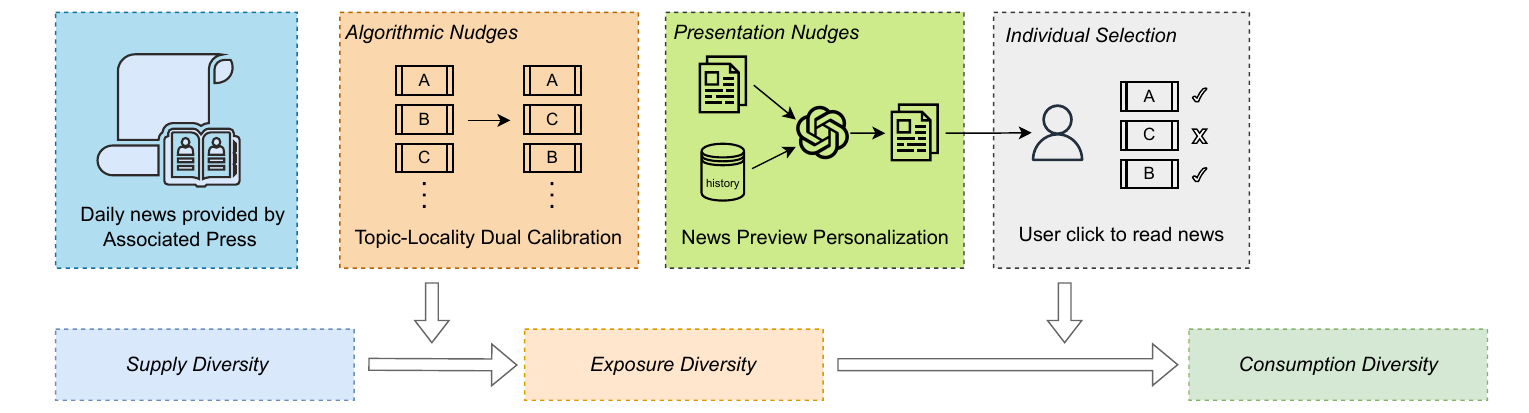} 
    \caption{Overview of our application of the personalized diversity nudge framework proposed by \citet{mattis_nudging_2024}.}
    \label{fig:teaser}
\end{teaserfigure}

\maketitle

\section{Introduction} 

In 2023, the Associated Press (AP) distributed over 375,000 stories to roughly 44 million monthly users~\cite{noauthor_ap_2024}. To manage this information overload, personalized news recommendation systems (RecSys) have become ubiquitous, leveraging past consumption behavior to filter content that matches the unique interests of users~\cite{li_survey_2019,okura_embedding-based_2017}. Although this approach successfully drives accuracy~\cite{zheng_drn_2018} and user engagement~\cite{liu_personalized_2010}, it runs the risk of creating feedback loops that reinforce narrow preferences~\cite{lorenz-spreen_how_2020}, generate unfair recommendations~\cite{wang_survey_2023}, and trap users in filter bubbles~\cite{liu_interaction_2021}. This creates a tension between the short-term benefits of personalization and the long-term societal goal of fostering a well-informed public.

To address this tension, prior work has categorized news diversity into three distinct stages: \textbf{supply diversity} (content available from publishers), \textbf{exposure diversity} (content visible to the user), and \textbf{consumption diversity} (content the user actively engages with)~\cite{van2011audiences, loecherbach2020unified, mattis_nudging_2024}. Since supply diversity is determined by editorial decisions, RecSys research has largely focused on optimizing \textit{exposure diversity} to mitigate bias~\cite{bauer_where_2024, sun2024interactive}. However, it is unclear how diverse recommendations can be converted into diverse consumption in practice. Empirical studies show that even when users are exposed to diverse viewpoints, they often ignore content that appears irrelevant or challenging to process~\cite{heitz_benefits_2022, levy_social_2021}. 
To bridge this divide, recent work has proposed a theoretical framework for personalized diversity nudges to not only show diverse content, but also facilitate consumption~\cite{mattis_nudging_2024, wu_personalized_2023}.



In this work, we build on \citet{mattis_nudging_2024}'s framework by empirically testing two novel nudging interventions: dual-calibration (an algorithmic nudge targeted at exposure) and LLM-generated presentation (a presentation nudge targeted at consumption). First, regarding exposure, we extend diversity beyond the traditional dimensions of politics and ideology~\cite{garimella_political_2021, ludwig_divided_2023} to explore news locality. While journalistic studies emphasize the importance of balancing global awareness with local relevance~\cite{castriota_national_2023, martin_local_2019}, RecSys research has rarely optimized for this balance. We propose a dual-calibration approach that balances both personalized \textbf{Topics} (e.g., Sports, Politics, Education, etc.) and diverse \textbf{Localities} (i.e., Domestic and World) to ensure U.S. users are exposed to both global and local perspectives relevant to them.

Second, to convert this exposure into consumption, we leverage Large Language Models (LLMs) to personalize news presentation. We rewrite news previews (i.e., headlines and subheads) to promote connections between articles recommended due to the locality component of dual-calibration and their prior news reading interest without altering the underlying facts. This approach aims to nudge users toward more balanced consumption by reducing the cognitive cost of engaging with unfamiliar topics or locations.

We evaluate the efficacy of this approach through a 5-week longitudinal study on the news recommendation platform POPROX. With that, we answer the following research questions:

\begin{description} 
    \item[\namedlabel{rq1}{RQ1}] How does \textbf{Locality + Topic Dual Calibration} affect exposure diversity compared to topic-only calibration strategy? 
    \item[\namedlabel{rq2}{RQ2}] How does Dual Calibration affect \textbf{consumption diversity} compared to topic-only calibration strategy? 
    \item[\namedlabel{rq3}{RQ3}] How do \textbf{LLM-generated presentation nudges} impact the conversion of exposure diversity into consumption diversity? 
    \item[\namedlabel{rq4}{RQ4}] How do \textbf{Dual Calibration + LLM nudges} impact user engagement and subjective satisfaction? 
\end{description}

Our contribution to this space is threefold: 

\begin{enumerate} 
    \item We present a novel diversity-aware RecSys that combines probabilistic dual-calibration (for algorithmic nudges) with LLM-based content reframing (for presentation nudges), and evaluated the system in a longitudinal experiment. 
    \item We empirically demonstrate the impact of dual-calibration for diverse consumption of domestic and global news, establishing \textbf{locality} as a valuable within-topic dimension of news diversity.
    \item We identify event-based narrative reframing as a promising direction for presentation nudges, suggesting a pathway to impact user selectivity.
    
\end{enumerate}

In the rest of the paper, we first discuss related work. We then introduce POPROX, our calibration and LLM personalization methods, followed by experiment design. We then share results from both offline training and an online randomized controlled field study. Finally, we discuss our findings, their limitations, and directions for future work. 

\section{Related Work}

\subsection{News Recommendation Diversity}

\subsubsection{Personalized Diversity}
Personalized diversity methods leverage users' historic behavior to align recommendations with their diverse interests to overcome pigeonholing in standard approaches like news recommendation approaches with multi-head self-attention (NRMS)~\cite{wu2019neural}.
Topic calibration ~\cite{abdollahpouri2020connection, steck_calibrated_2018, vrijenhoek2021recommenders} leverages proportions to align recommendation distributions with consumption patterns across topics.
Deep neural network approaches (i.e., heterogeneous graphs~\cite{zhang_heterogeneous_2024} and multi-view representation learning~\cite{chang_multi-view_2025}) enable richer personalization to users' dynamic preferences over time.

\subsubsection{Algorithmic Diversity Nudges}
Algorithmic diversity nudges intentionally skew recommendation distributions for diversity/fairness goals~\cite{wu_personalized_2023, hendrickx_news_2021, raza_news_2022}, introducing an accuracy trade-off~\cite{peng_reconciling_2024}.
Some have used topic-level distances to reward diverse exposure through taxonomy-based~\cite{rao_taxonomy_2013} and frequency measures~\cite {chakraborty2019optimizing} by regularizing recommendations.
Others target within-topic diversity measures (i.e., tone~\cite{tintarev_same_2018}, sentiment~\cite{wu_sentirec_2020}, and political leaning~\cite{heitz_benefits_2022}) to achieve similar goals.
Outside of news contexts, research has explored using LLMs for diversity-based re-ranking with mixed effects~\cite{carraro_enhancing_2025}.

Our work bridges these two approaches by combining topic calibration with non-personalized locality calibration to achieve personalized algorithmic diversity nudges.

\subsection{News Presentation Nudges}

News presentation nudges take many different forms and goals. 
Prior work uses pre- or post-recommendation nudges to promote socially responsible consumption via social comparison~\cite{agapie2015social} and self-reflection~\cite{munson2013encouraging}.
Others nudge to mitigate bias by pairing recommendations with alternative viewpoints for the same story~\cite{park2009newscube}.
Inspired by the persuasive success of LLMs for marketing~\cite{aldous2024using} and news summarization~\cite{huang-etal-2024-embrace}, recent work has focused on altering the news articles themselves to mitigate bias~\cite{ruan2024rewriting, raza2024dbias}, depolarize content~\cite{liu2021political}, and apply positive/negative reframing~\cite{10.1145/3640457.3688008}.
Closest to our work, \citet{gao_generative_2024} pair generative news recommendation with theme- and semantic-level representations to create personalized narratives between recommended articles.

Together, many of these algorithmic and presentation nudging approaches demonstrate success in offline evaluations.
However, few conduct user studies--and even fewer evaluate the effects of these nudging interventions in realistic contexts~\cite{mattis_nudging_2024}.
Our results build on this limited understanding of the longitudinal effects of diversity-aware news RecSys~\cite{10.1145/3640457.3688008}, which is particularly relevant given prior work suggesting news diversity benefits long-term platform growth~\cite{wang_survey_2023, liu_interaction_2021}.

\section{POPROX Platform}
\label{poprox}
We ran this study on POPROX \cite{burke2024conducting}, an experimental platform maintaining U.S. news readers and distributing daily personalized newsletters from Associated Press (AP) to them. Researchers can design different algorithms or interfaces to change users' newsletter content and evaluate their effectiveness with interaction log and survey data. POPROX also provides a suite of base recommender algorithms and APIs for experimenters to integrate and modify. 

\textit{Internal Pilots.} We conducted two sets of internal pilots to refine our algorithm design and verify system functionality before the experiment\footnote{We appreciate support from POPROX researchers.}.
During development, we conducted an \textit{ongoing pilot} with 24 internal users provided by POPROX, where code changes were pushed to users' daily POPROX newsletters.
Here, we improved \ref{sec:method-calibration} and \ref{sec:context-generation} through iterative feedback that informed threshold and prompt tuning.
Following these refinements, we ran a \textit{dark live} trial for a week before the actual experiment launch with 8 users per treatment group to verify experiment infrastructure. 

\section{Method} 
\label{Method}
\begin{figure*}[!htbp]
\centering
  \includegraphics[width=2.0\columnwidth]{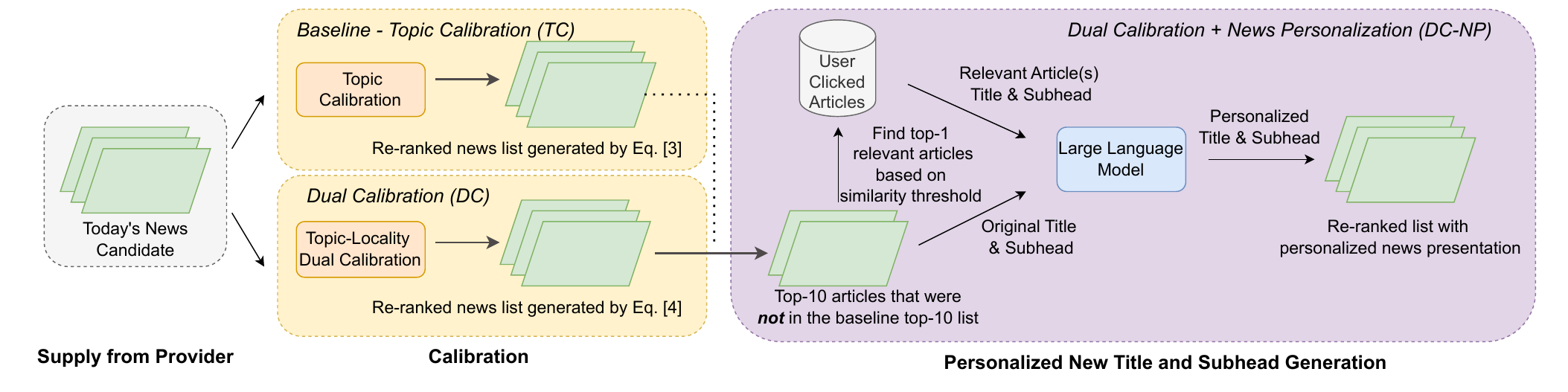}
  \caption{Method overview of the calibration and personalized news preview generation workflow.}~\label{fig:method-overview}
\end{figure*}

The general method can be split into two parts: 1) recommendation re-ranking based on Topic Calibration (TC) and Topic-Locality Dual Calibration (DC); and 2) news preview personalization of re-ranked news from DC treatment that would not have been recommended in the baseline TC algorithm (DC-SC) (see Fig. \ref{fig:method-overview}). The following subsections provide detailed explanations of each component. Before diving into the details, we introduce key notation: Let $A = \{a_1\dots,a_n\}$ represent the set of articles the news provider released today, and $H$ represent the set of articles a user has clicked on previously. Let $I_{10}$ and $J_{10}$ denote the set of top-10 articles re-ranked by TC and DC, respectively. Personalized news previews are applied to articles that are in $I$ but not in $J$, and we define this set as $S = \{s_1, \dots, s_k\}$. 





\subsection{Calibration} \label{sec:method-calibration}
To align a body of recommended items with a user's consumption preferences along a specific axis, we draw inspiration from the calibration framework formalized by \citet{steck_calibrated_2018}.
For news, topic calibration suggests that if a user has clicked on 20 politics and 80 sports articles, their personalized newsletter should consist of about 20\% politics and 80\% sports articles. 
However, prior work has highlighted the risks of this approach, particularly its potential to reinforce existing user biases. 
In response, researchers have proposed various dual-optimization strategies to balance calibrated recommendations with broader system objectives~\cite{wang_two-sided_2023, zhang_evolution_2023}.

Following prior work~\cite{steck_calibrated_2018, oh_novel_2011}, we adopt Kullback-Leibler (KL) divergence as a calibration metric to re-rank recommendations along two dimensions: topic (see Eq.~\ref{eq:topic_divergence}) and locality (see Eq.~\ref{eq:locality_divergence}).
Let $p(t \mid u)$ denote the distribution of topics $t$ that user $u$ has previously clicked on.
Similarly, $q(t \mid u)$ denotes the topic distribution within a user's newsletter.
Let $p(l \mid u)$ denote the distribution of localities $l$ in articles released by news provider today and $q(l \mid u)$ denote the locality distribution within a user's newsletter. 
\begin{equation}
C_{KL}^{t}(p, q)=KL_t(p \,\|\, q) = \sum_{t} p(t \mid u) \log \frac{p(t \mid u)}{q(t \mid u)}
\label{eq:topic_divergence}
\end{equation}
\begin{equation}
C_{KL}^{l}(p, q)=KL_l(p \,\|\, q) = \sum_{l} p(l \mid u) \log \frac{p(l \mid u)}{q(l \mid u)}
\label{eq:locality_divergence}
\end{equation}


The recommended newsletters $I_{10}$ and $J_{10}$ are iteratively populated with today's articles by optimizing Eq.~\ref{eq:topic_calibration} and Eq.~\ref{eq:dual_calibration} respectively:
\begin{equation}
i_{n+1} = \arg\max_{i \in A \setminus I_n} (1 - \theta_l - \theta_t)S_{nrms}(i) - \theta_t C_{KL}^{t}(p, \tilde{q}) 
\label{eq:topic_calibration}
\end{equation}
\begin{equation}
j_{n+1} = \arg\max_{j \in A \setminus J_n} (1 - \theta_l - \theta_t)S_{nrms}(j) - \theta_l C_{KL}^{l}(p, \tilde{q}) - \theta_t C_{KL}^{t}(p, \tilde{q}) 
\label{eq:dual_calibration}
\end{equation}
with $\theta_l, \theta_t \in [0, 1]$ and $S_{nrms}(I) = \sum_{i \in I} s_i$ representing the predicted user preference scores for the newsletters in $I$, as provided by a base NRMS~\cite{wu2019neural}.

For topic calibration, our implementation leverages 14 existing high-level topics provided by POPROX that were derived from AP's subject metadata for user preference modeling.\footnote{POPROX's list of high-level news topic tags is adapted from a subset of AP's broad subject metadata~\cite{noauthor_ap_nodate} and includes: \textit{U.S. news, World news, Politics, Business, Entertainment, Sports, Health, Science, Technology, Lifestyle, Religion, Climate and environment, Education, Oddities}.}
For locality calibration, we further adapt AP's subject metadata to define an article's locality distribution: \textbf{Domestic}, \textbf{World}, and \textbf{Neither}.
Specifically, we leverage AP's \textit{General news} categories to define Domestic news with the AP tags \textit{U.S. news} or \textit{Washington news}, World news with the AP tag \textit{World news}, and Neither if it doesn't meet either categorization.
\footnote{AP applies the tag \textit{Washington news} instead of \textit{U.S. news} for local Washington stories of national interest~\cite{noauthor_ap_nodate}. We consolidate both tags to avoid these stories being classified as Neither. In practice. Neither often relates to general news stories without a geographic affinity.}
Articles often have multiple AP subject tags, meaning they can have multiple topics or be classified as both Domestic and World localities. 
We acknowledge this as a provider choice for the nature of AP news content. 
This quality is also what positions locality as a within-topic diversity goal, as U.S., Washington, or World news tags often intersect with other subject tags.

\subsubsection{Calibration Tuning}
To efficiently find the optimal calibration terms for the topic ($\theta_t$) and locality ($\theta_l$) pair, we employed random search \cite{bergstra2012random}. The search space was defined as the Cartesian product of $\theta_t$ and $\theta_l$ values in the range of [0,1] with a 0.05 step. Following \cite{bergstra2012random}, we randomly sampled 60 configurations, which provides a 95\% probability of identifying a solution within the top 5\% of the search space. The optimal pair ($\theta_t^*, \theta_l^*$) was selected to maximize a utility score $S$. This objective function rewards high ranking quality (normalized $nDCG@10$) while penalizing calibration divergence (normalized $C_{KL}^{t}$ and $C_{KL}^{l}$), weighted by $(0.4, 0.3, 0.3)$ respectively:

    \begin{equation} 
        \label{eq:combined-score} S(\theta_t, \theta_l) = w_{nDCG} \cdot \widetilde{nDCG} - w_{C_{KL}^{t}} \cdot \widetilde{C}_{KL}^{t} - w{C_{KL}^{l}} \cdot \widetilde{C}_{KL}^{l}
    \end{equation} 
    \begin{equation} 
        \label{eq:theta-definition} (\theta_t^*, \theta_l^*) = \operatorname*{arg,max}{(\theta_t, \theta_l) \in \Omega} S(\theta_t, \theta_l) 
    \end{equation}





\subsection{Personalized Preview Generation} \label{sec:context-generation} 

\begin{figure*}[t]
\centering
  \includegraphics[width=\textwidth]{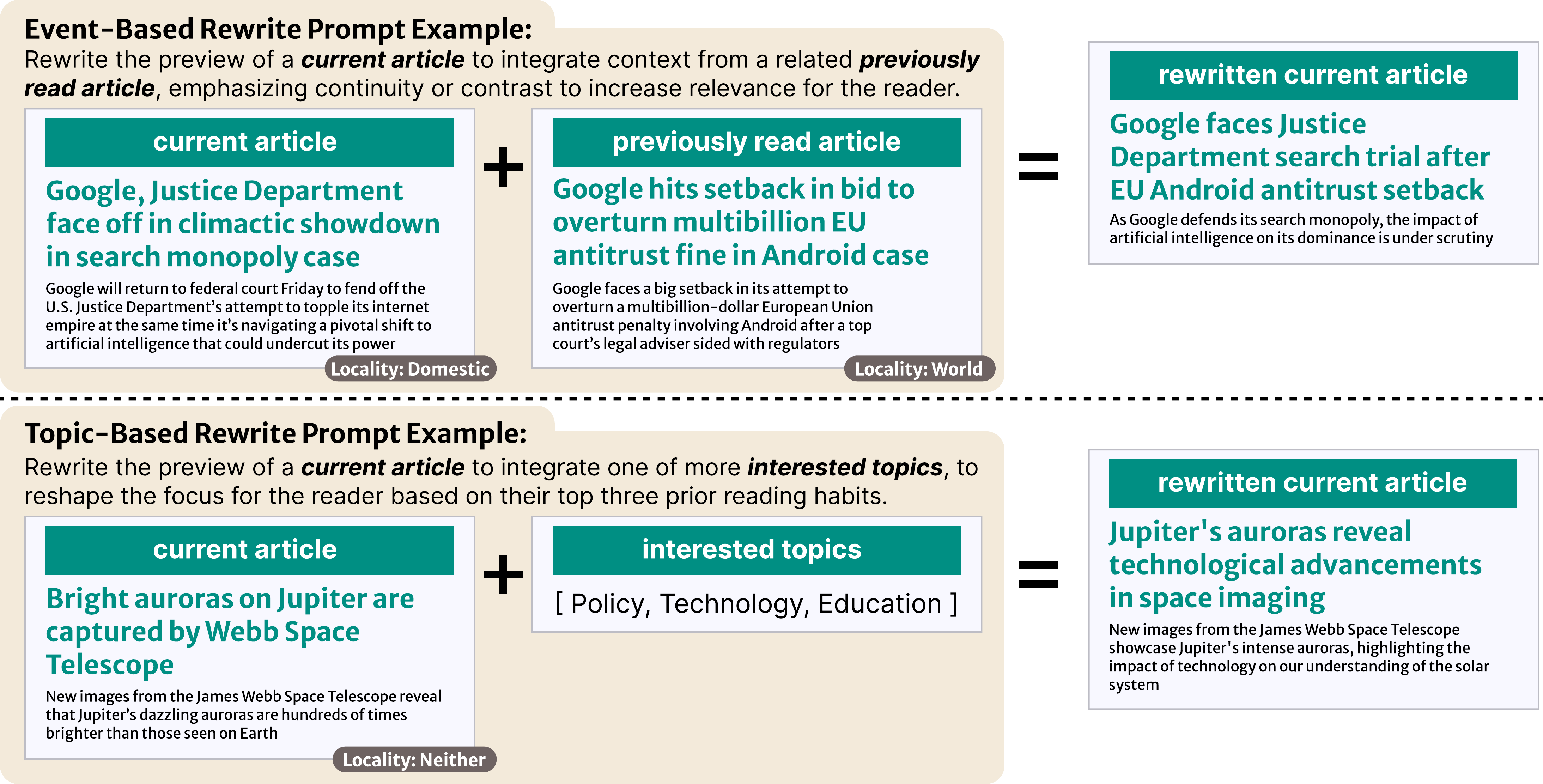}
  \caption{Examples of rewrite strategies. Event-based prompts (top) integrate context from a previously clicked article with high cosine similarity. Topic-based prompts (bottom) fall back to incorporating top-3 interested topics. Prompt language shown is illustrative. Detailed prompts are included in Appendix Figure~\ref{fig:detail-prompt}.}~\label{fig:prompt}
  \label{fig:prompt_example}
\end{figure*}

As illustrated in Figure~\ref{fig:teaser} and~\ref{fig:method-overview}, we generated personalized news previews on re-ranked news to understand how this presentation nudges impacts individual selection. 
Drawing on the concept of self-actualization nudges~\cite{knijnenburg2016recommender, mattis_nudging_2024}, our approach aims to help users discover news beyond their established preferences by surfacing relevant connections between unfamiliar articles added as a result of the locality component of DC and their prior reading history or stated interests.
Our goal is to reduce the cognitive load and attention barriers that have limited the effectiveness of prior self-actualization interventions~\cite{liao2014can} by embedding explanations of relevance directly into news headlines and subheads.
The process consists of two steps: 1) identifying relevant context articles, and 2) generating the new preview with an LLM using both the original news article and relevant context. The purpose of identifying relevant context articles from $H$ is to ensure that the generated news preview is more closely aligned with the user's topic preferences or relates to an event the user has previously read. 

To measure relevance, we apply sentence transformer embeddings to compute cosine similarity. For each article $s_i \in S$, we check if there exists an article $h_j \in H$ such that the cosine similarity between their embeddings exceeds a tuned threshold $\theta$ (see Sec.~\ref{sec:threshold_tuning}). This would lead to two possible rewrite strategies: 

\begin{enumerate}
    \item \textbf{Event-based preview:} when there exists at least one article $h_j \in H$ with a cosine similarity higher than $\theta_{sim}$.
    \item \textbf{Topic-based preview:} when there is no article from $H$ has a cosine similarity higher than $\theta_{sim}$.
\end{enumerate}

For event-based previews, we prompted GPT-4o-mini to rewrite the headline and subhead by considering the connection between $s_i$ and the article $h_j \in H$ with the highest similarity. For topic-based rewrites, we selected the top-3 topics of users based on their POPROX onboarding reading interest survey and their previously clicked articles in $H$. We then prompt the LLM to rewrite the headline and subhead by incorporating these high-level topic preferences (see Fig.~\ref{fig:prompt_example} for an example of each rewrite strategy).

Similar to prior work~\cite{zent_piivot_2025}, we employ feedback-based reprompting to enforce measurable qualities of the news previews. 
To meet the POPROX API latency requirements, rewrite operations were executed in parallel. 
Feedback is structured in two categories: 1) article-level feedback that enforces style guidelines (e.g., AP capitalization rules), and 2) newsletter-level feedback that ensures semantic diversity across headlines within a single newsletter.


\subsubsection{Threshold Tuning}
\label{sec:threshold_tuning}


To determine the optimal similarity threshold $\theta_{sim}$, we combine quantitative tuning and internal user pilots. 
Where prior work has used LLMs to evaluate threshold/coherence tradeoffs offline for generative news narratives~\cite{gao_generative_2024}, our method aims to maximize opportunities for legitimate event-level rewrites while maintaining coherence and trust using summary quality metrics and human evaluation. 
We use ROUGE-L~\cite{lin-2004-rouge} for initial parameter tuning and acknowledge its limitations as an imperfect proxy metric for rewrite quality, which we address through ongoing internal pilot testing.\footnote{ROUGE-L measures the similarity between the rewritten and original previews, making it useful for ensuring the rewrite accurately represents the original article. However, it does not account for our objective of incorporating informative context from another source that reduces similarity.}
Rounds of offline testing showed that higher similarity thresholds exhibit diminishing returns in summary quality at $\theta_{sim} = 0.3$ for event-level rewrites.
Therefore, we initialize $\theta_{sim}$ here to capture this inflection point for pilot evaluation.
During internal pilots, users provided informal feedback identifying both successful and problematic rewrites. 
For example, one flagged example inappropriately compared an article about Virginia Giuffre's death to a prior article about 7-Eleven leadership changes, framing both as stories of ``determination.'' 
Analysis of this user feedback alongside article metadata revealed that $\theta_{sim} = 0.4$ cleanly separated inappropriate examples from positive rewrites, leading us to adopt this more conservative threshold during the experiment.


\section{Experiment Design} 

In this section, we detail the experimental design of a randomized controlled trial of our method (Sec.~\ref{Method}).
Our study lasted five weeks, from May 18, 2025, until June 22, 2025. 
In total, 120 active U.S. POPROX readers were assigned to the experiment, with 40 participants randomly distributed to each of the three treatment groups. 
These treatments directly impacted the top-10 news articles of users' daily newsletter during the experiment period.

\subsection{Treatment Groups}

We designed three treatment conditions to understand the effects of dual calibration and personalized preview generation in relation to our topic calibration baseline.
Each treatment group builds upon the prior treatment definition to control for algorithmic confounds and isolate the effects of the added component.

\begin{enumerate}
    
    \item  \textit{\textbf{(Baseline) Topic Calibration (TC)}}: Applied topic calibration based on Eq.~\ref{eq:topic_calibration} to the base NRMS model.

    \item  \textit{\textbf{Dual Calibration (DC)}}: Applied additional locality calibration to the baseline TC based on Eq.~\ref{eq:dual_calibration}.
    
    \item  \textit{\textbf{Dual Calibration + News Personalization (DC-NP)}}: Applied personalized news previews to DC based on Sec. \ref{sec:context-generation}.
    
\end{enumerate}

\subsection{Voluntary Survey Response} 

To collect subjective user reading experience data, POPROX distributes surveys to its readers in a 5-week cycle. Subscribers receive a survey email about diversity, accuracy, satisfaction, control, and other perceptions of their past week's reading experience. We also designed and distributed study-specific pre- and post-questionnaires for participants to fill in voluntarily. Full survey questions are available in Appendix Table \ref{tab:survey_questions}. All survey results are presented as descriptive findings and do not represent statistical tests.


\subsection{Analytical Methods}

To isolate the effects of our experiment manipulations, we fit generalized linear mixed-effects models (GLMMs) using the \texttt{lme4} package in R.
We use articles clicked in each newsletter as our primary signal of interest.
This choice was constrained by the functionality available in the POPROX platform and ethical implications of further tracking; we discuss the limitations of this heuristic further in Section~\ref{sec:limitations}.
We include data from 67 participants who clicked on at least one article. 
For each model, we z-score all continuous predictors to improve modeling performance and stability. 
We use the BOBYQA optimizer with an increased maximum function evaluation limit of 200,000 to ensure model convergence given the inclusion of interaction terms.
We include predictor variables based on our RQs, theoretical interest, availability within platform constraints, and nested model testing.
These modeling decisions allow us to control for and understand individual, temporal, and algorithm design differences.

For exposure diversity and consumption diversity, we fit two gamma models with log links to predict KL $ExposureDivergence_j$ and $ConsumptionDivergence_j$ of all recommended articles up to and including newsletter $j$ from AP's supply diversity during the experiment.
We use KL divergence because it is explicitly optimized for in our calibration model.
Because this metric is asymmetric, we preserve the same directionality throughout calibration and evaluation, treating the AP supply distribution as the reference and the user’s true exposure or consumption distribution as the target (i.e., the information lost when approximating AP supply distribution using observed exposure or consumption).\footnote{Because this is scale-invariant, we scale divergence metrics for numerical stability modeling and rescale intercepts to report coefficients in their original scale.}
Similar to \citet{10.1145/3610071}, we report the formulas for these models below, where $a_{k[j]}$ is the random intercept that can vary for each user $k$, $Group_{j}$ is the experiment group, $Week_{j}$ is the experiment week newsletter $j$ was sent, and $\beta$ (omitted for space) are the estimated coefficients for each predictor variable.
We use nested model testing to validate factors for Equation~\ref{eqn:formula1} and~\ref{eqn:formula2}.\footnote{Nested likelihood ratio tests for Eq.~\ref{eqn:formula1} start with o $Group_j$ and sequentially add: $a_{k[j]}$ (chisq(2)=835.4***), $Week_i$ (chisq(1)=835.4***), and $Group_i x Week_i$ (chisq(2)=74.3***). For Equation~\ref{eqn:formula2}: $a_{k[j]}$ (chisq(1)=823.2***), $Week_i$ (chisq(1)=823.2***), and $Group_i x Week_i$ (chisq(2)=12.6**). ***p<.001, **p<.01.}

\begin{align}
\log(Exposure&Divergence_j) = \label{eqn:formula1}\\&a_{k[j]} + Group_j + Week_j + Group_j:Week_j \notag\\[4pt]
\log(Consumption&Divergence_j) = \label{eqn:formula2} \\&a_{k[j]} + Group_j + Week_j + Group_j:Week_j \notag
\end{align}

For engagement, we fit two binomial models with a logit link to predict whether each recommended article $i$ was clicked ($Clicked_i$).
Let $Position_i$ denote the rank of $i$ in the newsletter. 
We define $PriorTopicInterest_i$ and $PriorLocalityInterest_i$ as the cosine similarity of $i$'s topic and locality vectors to $k$'s historic clicks.
Finally, let $PromptLevel_i$ denote the prompting strategy used to present article $i$.
We use nested model testing to validate factors for Equation~\ref{eqn:formula3} and retain this structure for Equation~\ref{eqn:formula4}.\footnote{Nested likelihood ratio tests for Eq.~\ref{eqn:formula3} start with $Group_i$ and sequentially add: $a_{k[i]}$ (chisq(1)=1300.4***), $Position_i$ (chisq(1)=145.3***), $Week_i$ (chisq(1)=56.5***), $PriorTopicInterest_i$ (chisq(1)=149.9***), $PriorLocalityInterest_i$ (chisq(1)=6.5*), and $Group_i x PriorTopicInterest_i$ (chisq(2)=12.9**).  ***p<.001, **p<.01, and *p<.05.}
Note that $PromptLevel_i$ is only available for the DC+NP group. Therefore, we fit a separate model (Eq.~\ref{eqn:formula4}) to understand its effects.

\begin{align}
\logit(P(Clicke&d_i = 1)) = \label{eqn:formula3} \\ a_{k[i]} &+ Group_i + Position_i + Week_i \notag \\
   &+ PriorTopicInterest_i + PriorLocalityInterest_i \notag \\
   &+ Group_i{:}PriorTopicInterest_i \notag \\[6pt]
\logit(P(Clicke&d_i = 1 \mid Group_i = \text{Context})) = \label{eqn:formula4} \\ a_{k[i]} &+ PromptLevel_i + Position_i + Week_i \notag \\
   &+ PriorTopicInterest_i + PriorLocalityInterest_i \notag \\
   & + PromptLevel_i{:}PriorTopicInterest_i \notag
\end{align}

\subsection{Ethical Consideration}

\quad \textit{Institutional Review. } The POPROX experiment platform was reviewed and approved by a third-party ethical review board ADVARRA \footnote{https://www.advarra.com/}, and our specific study design were reviewed and approved by our local institute review board (IRB).

\textit{Generative AI Usage.} We used ChatGPT, Claude, and Gemini to support code development, debugging, and edit grammar and styling when polishing the paper draft after initial designs and text were produced by the authors. All generated content was reviewed by the authors for correctness and represents our original ideas. 

\textit{Positionality.} Our research team is composed of both internal and external members of the POPROX research platform. As a new system, this afforded us privileged access to support from POPROX as we piloted its experimental infrastructure. We are conscious of this positionality throughout our research and prioritized maintaining these boundaries when possible to support POPROX's evolving processes for engaging with external researchers.



\section{Results} 
\label{sec:results}


Across 120 participants equally distributed among three treatment groups, we collected 1697 article clicks from 915 personalized newsletters on 633 unique news articles. We also distributed seven surveys to participants to voluntarily fill in, including five regular weekly POPROX surveys, and two customized study surveys to get pre- and post-survey perception on the dual-calibration and news rewrite. In total, we collected 94 surveys from 28 participants. Due to the limited number of survey responses to reach statistical significance, we primarily utilized survey results to identify observational trends that complement our statistical analysis and inform future work.

\subsection{Training/Offline Results}

\begin{figure*}[!htbp]
\centering
  \includegraphics[width=1.8\columnwidth]{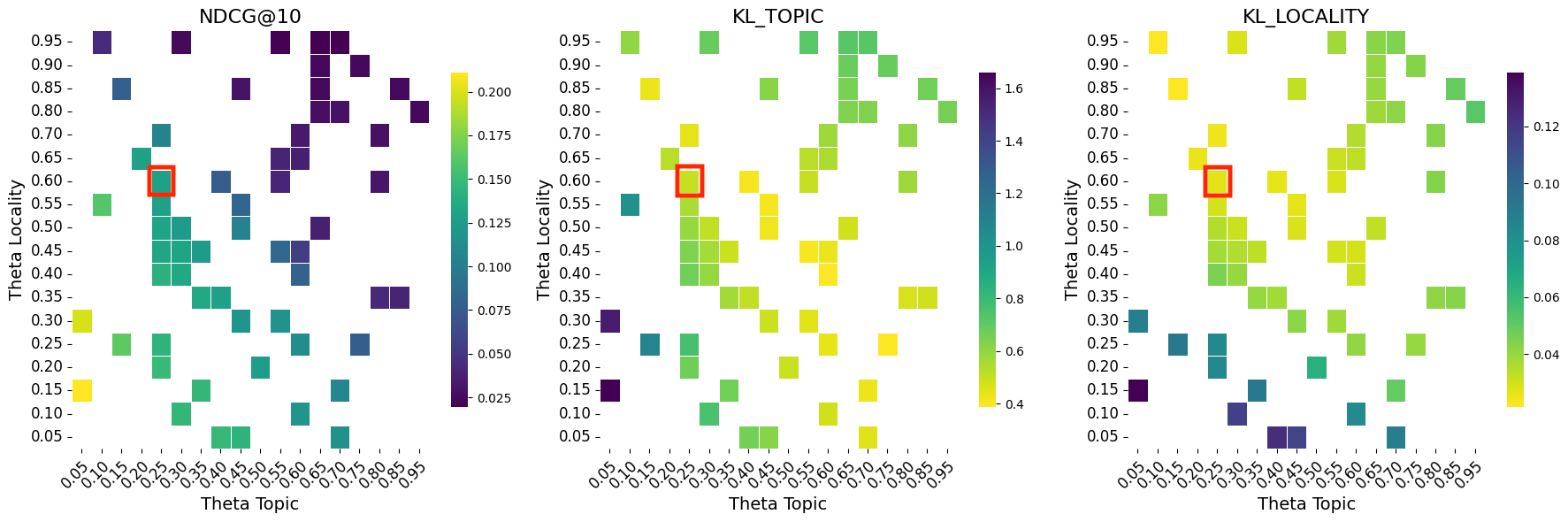}
  \caption{Theta pair tuning metrics based on the one-month prior-experiment evaluation data. A higher NDCG@10 score is preferred, and a lower KL divergence score on theta topic and theta locality indicates better alignment. The red boxes indicate the final theta pair selection based on Eq. \ref{eq:theta-definition}.}~\label{fig:theta-tuning}
\end{figure*}

We trained the three recommendation models based on historic data of participants and tuned the theta values for \textit{DC} and \textit{DC-NP} models following the calibration methods described in Section~\ref{sec:method-calibration} and~\ref{sec:threshold_tuning}. Specifically, we collected four weeks of news reading recommendation and activity data from participants before the actual experiment, using the first three weeks for training, and the last week for evaluation. The heatmap for the 60 randomly sampled theta pairs on original ($nDCG@10$, $C_{KL}^{t}$, $C_{KL}^{l}$) values on the evaluation data is presented in Figure~\ref{fig:theta-tuning}. Based on normalized final score calculation with heuristic weights (Eq. \ref{eq:theta-definition}), the final theta values for $\theta_l$ and $\theta_t$ is set as $(0.6, 0.25)$. 

\subsection{Exposure Diversity (RQ1)}

\begin{figure}[t!]
    \centering
    \includegraphics[width=3in, keepaspectratio]{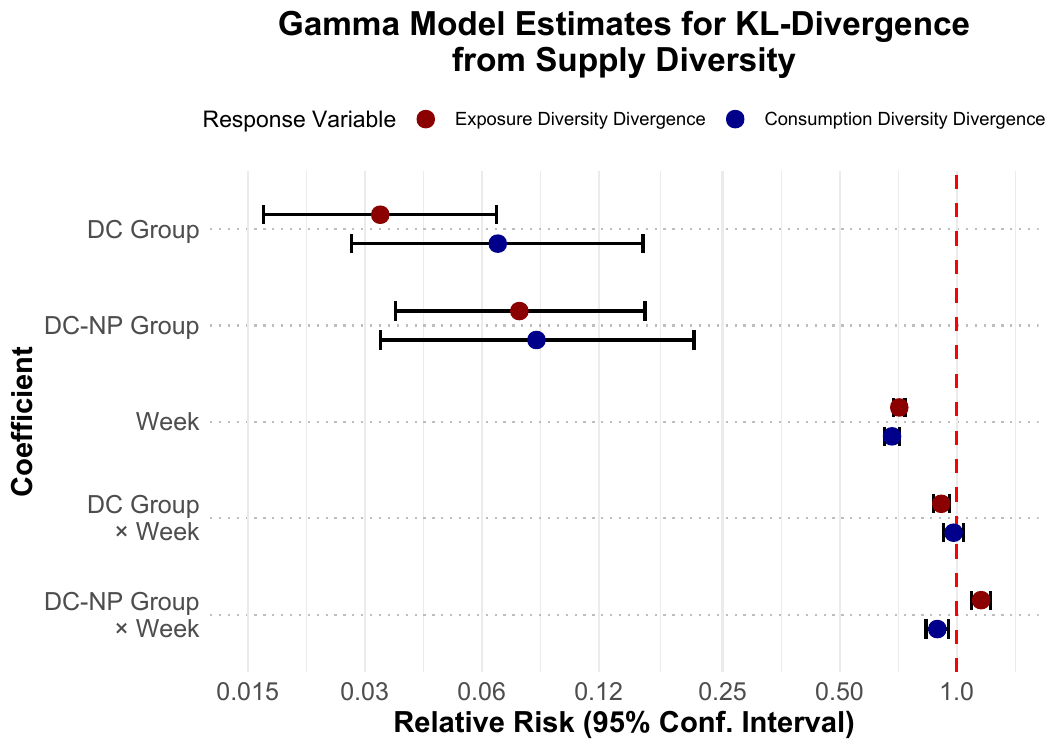}
        \caption{Forest plots of relative risk (log scale) from Eq.~\ref{eqn:formula1} (\textcolor{darkred}{red}) and Eq.~\ref{eqn:formula2} (\textcolor{darkblue}{blue}). Smaller is better. Group comparisons relative to TC baseline with moderate KL-divergence (exposure intercept: 0.06 [0.04, 0.10], consumption intercept: 0.10 [0.06, 0.19]). Both treatments significantly reduce exposure divergence (DC: RR=0.03, p<.001; DC-NP: RR=0.08, p<.001) and consumption divergence (DC: RR=0.07, p<.001; DC-NP: RR=0.08, p<.001). Notably, there is no significant difference between DC and DC-NP consumption divergence. Week had a small, significant effect reducing exposure diversity  (RR=0.71, p<.001) and consumption diversity (RR=0.68, p<.001) for the TC group, with minor deviation of mixed significance by treatment interactions.}
    \label{fig:divergence}
\end{figure}


We evaluated the impact of algorithmic nudges on exposure diversity by comparing the DC and DC-NP treatment groups to the TC baseline; recall that both treatment groups received the same underlying algorithmic nudges.
As shown in Figure~\ref{fig:divergence}, both treatments were associated with markedly lower KL exposure divergence relative to the TC baseline: the DC treatment experienced a reduction $\sim$97\% and the DC-NP treatment $\sim$92\%. 
Divergence decreased over time for the TC group ($\sim$29\% reduction per standardized week). 
From survey responses, we observed that users' perceptions of the importance of locality also shifted for different treatment groups. Both treatment groups have unchanged agreement percentage for their pre- and post-response on the question \textit{"I feel that it is important that a newsletter provides a variety of international and domestic news topics to its readers."} while the control group agreement reduced $\sim$14\% from pre- to post survey. 


\textbf{RQ1. Locality+topic dual-calibration successfully increased exposure diversity compared to topic-calibration alone, with treatment effects cutting divergence from supply diversity distributions by nearly 2x.} 
Primarily, these improvements were achieved across users with varying topic preferences, demonstrating the practical feasibility of optimizing across these calibration dimensions for our experiment population.

\subsection{Consumption Diversity}

\subsubsection{Topic-Locality Algorithmic Nudges (RQ2)}


Next, we evaluated the impact of algorithmic nudges on consumption diversity by comparing the DC group to our TC baseline.
Figure~\ref{fig:divergence} shows that DC was associated with large reductions in locality consumption divergence ($\sim$93\%) relative to TC.
This divergence also decreased over time for the TC group ($\sim$32\% reduction per standardized week).
In the post survey, DC users agreed $\sim$33\% more than TC group users that they \textit{"would like the future POPROX newsletters to also balance international and domestic news."}

\textbf{RQ2. Locality+topic dual-calibration also increased consumption diversity compared to topic-calibration alone, with a similar effect size to exposure diversity effects.}

\subsubsection{Personalized Presentation Nudges (RQ3)}

Next, we observed the joint effect of dual calibration and personalized news previews by comparing the DC-NP group to our TC baseline.
Figure~\ref{fig:divergence} shows that DC-NP treatment was associated with large reductions in locality consumption divergence ($\sim$92\%) relative to the TC group, but we find no significant differences between DC and DC-NP. 
In the post survey, participants in the DC-NP treatment group expressed higher agreement compared to the control group (16.67\%) that increased locality balance is an improvement over POPROX's baseline algorithm (\textit{"This balance was intended to recommend a similar ratio of international and domestic news based on our source provider. Reflecting on the balance, do you think that's an improvement?"}). They also agreed $\sim$8\% more than TC group users that they would like future newsletters to balance international and domestic news. 
\textbf{RQ3. We find that adding LLM-driven personalized previews to dual calibration does not significantly improve consumption diversity beyond the effects of algorithmic nudges alone.}
In the following section, we further reflect on the impact of our implementation on engagement to identify potential design refinements and support future work on personalized news presentation.

\subsection{Engagement (RQ4)}
\label{sub:engagment}

\subsubsection{Topic-Locality Algorithmic Nudges}


We used our click likelihood logistic regression model to understand the effects our experimental design had on engagement.
Figure \ref{fig:engagement} shows that differences in group-level effects were not significant; any group differences were primarily explained by user-level variance.
Prior topic and locality interest increased click odds by 22\% and 8\% per SD, respectively.
Notably, interactions between treatment and topic interest were significant, meaning the effect of prior topic interest was stronger for users in DC and DC-NP treatment groups by ~28\% and ~21\%, respectively.
\textbf{RQ4. Reflecting on the TC and DC components, we find users in our baseline had ~4\% chance of clicking a recommended article, controlling for position, experiment week, and topic + locality interest; we find no evidence to suggest the unpersonalized locality component of the DC treatment negatively impacted overall engagement.}

\subsubsection{Personalized Presentation Nudges}
\label{sec:presentation_nudges}

\begin{figure*}[t]
    \centering
    \textbf{Binomial Model Estimates for Click Likelihood}\par

    \begin{subfigure}[t]{0.48\textwidth}
        \centering
        \textit{By treatment group}\par\vspace{0.5em}
        \includegraphics[width=\linewidth]{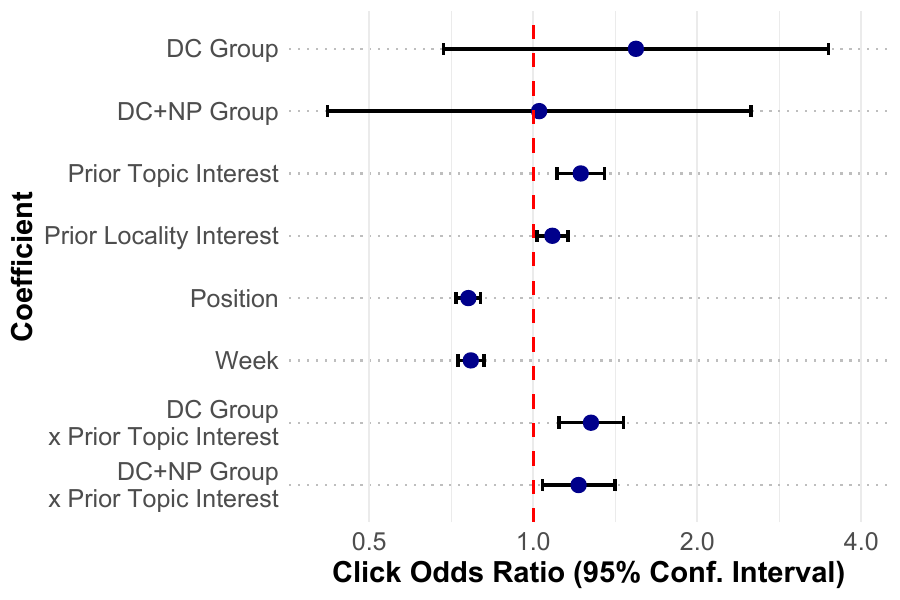}
        \caption{Forest plots of odds ratios (log scale) from Eq.~\ref{eqn:formula3}. Group comparisons relative to TC baseline (intercept: 0.04 [0.02, 0.07]). Neither DC (OR:1.5, p=.36) nor DC-NP (OR:1.03, p=.36) increased overall click likelihood , but both significantly amplified the effect of prior topic interest (DC×Topic: OR=1.28, p<.001; DC-NP×Topic: OR=1.21, p<.01), indicating that calibration increased selectivity toward familiar topics. Position (OR=0.76, p<.001) and week (OR=0.77, p<.001) had a strong negative effect on engagement.}
        \label{fig:engagement}
    \end{subfigure}
    \hfill
    \begin{subfigure}[t]{0.48\textwidth}
        \centering
        \textit{By rewrite strategy}\par\vspace{0.5em}
        \includegraphics[width=\linewidth]{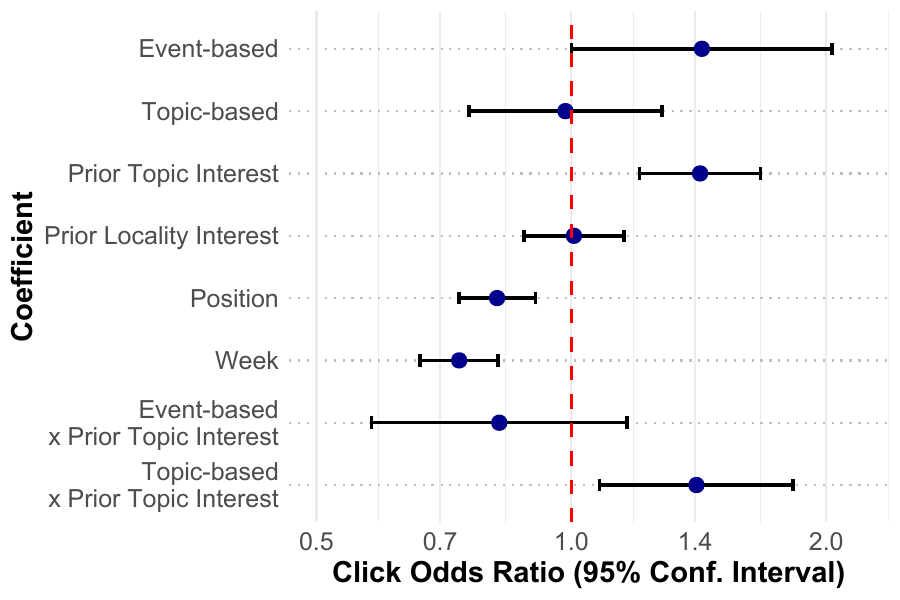}
        \caption{Forest plots of odds ratios (log scale) from Eq.~\ref{eqn:formula4}. Rewrite comparisons relative to no rewrite (intercept: 0.04 [0.02, 0.07]). Event-based rewrites significantly increased clicks (OR=1.42, p<.05), controlling for prior topic interest (OR=1.42, p<.001) and position (OR=0.82, p<.001). Topic-based rewrites amplified the effect of prior interest (OR=1.41, p<.01), while event prompts did not (OR=.82, p=.48), suggesting event-based rewrites can impact user selectivity with more statistical power.}
        \label{fig:prompt_engagement}
    \end{subfigure}

    \label{fig:click_models_combined}
\end{figure*}

Finally, we evaluate the impact of DC-NP and the rewrite strategy used to personalize each article preview on engagement.  
During the experiment, the algorithm applied varying levels of personalization across 5,362 total articles: 531 articles (10\%) received event-based rewrites, 1,711 articles (32\%) received topic-based rewrites, and 3,120 articles (58\%) received no rewrite.
Similar to \ref{sub:engagment}, we find no significant effects of DC-NP on overall click engagement in Figure \ref{fig:engagement}.

Within the DC-NP treatment group, Figure \ref{fig:prompt_engagement} shows that most prompt-level differences are explained by user-level variance, but event-level prompts showed a large, marginally significant increase (p<0.05) in click odds by 43\%.
User-level topic interest strongly predicted clicks, while position and week also had significant negative effects.
We find a marginal interaction effect between topic prompts and users' prior topic interest, meaning that the effect of prior interest was more salient for articles with topic-level rewrites by $\sim$40\%.

\textbf{RQ4. Reviewing the DC-NP component, we find no evidence to suggest that personalized preview generation had a negative impact on engagement.}
Moreover, analyzing engagement by rewrite strategy reveals a more nuanced perspective: while the more specific event-based prompts were only applied 10\% of the time due to safety and legitimacy constraints, they were effective in promoting clicks even when controlling for known algorithmic confounds like prior topic interest. 
Interestingly, the post-survey response showcased a trend that users in DC-NP felt they had more control over their newsletter (16.67\%) compared to both DC and TC groups (i.e., \textit{"I feel like I have control over what articles POPROX News selects."}).

\section{Discussion}

Here, we reflect on our experiment results through the lens of prior work and the nudging for diversity framework~\cite{mattis_nudges_2025}.
We examine the mechanisms of our algorithmic and presentation nudges, as well as broader implications for designing for diverse news RecSys.

\subsection{Nudging for Diversity}
Our results both align with and diverge from prior work on news diversity in meaningful ways.
As with prior work~\cite{jurgens_mapping_2022}, our findings suggest a close relationship between exposure and consumption diversity when diverse content is personalized to individual preference.
We observed that calibrating users' exposure diversity to our goals had an immediate effect on consumption diversity; dual-calibration (DC) reduced locality consumption divergence by $\sim$93\% relative to the control.
Moreover, nested model testing for Figure~\ref{fig:engagement} revealed no significant interactions between week and our treatment conditions compared to our baseline, suggesting this improvement didn't come at the cost of longer-term engagement, as echoed by some recent findings in news recommenders \cite{yu2024nudging,mattis_nudges_2025}. 
Notably, the TC baseline improved over time with respect to locality diversity without any manipulations.
This pattern is distinct from discussions of political news consumption on social media~\cite{levy_social_2021}, as users naturally explored domestic and global news in response to journalism cycles.
While the immediate responsiveness and uniform behavior shifts of dual-calibration may be important for certain objectives, our results emphasize that one-time user studies may overestimate the effect sizes of diversity-aware RecSys.
Echoing prior calls~\cite{jesse_digital_2021}, we highlight the need to continue to understand the longer-term effects of nudging for news diversity.

Data-driven nudges have received considerable attention recently.~\cite{sadeghian_data-driven_2024, shin_value_2024, shin_engineering_2025}
Reflecting on the nudging framework, the lack of significant differences between our treatment groups warrants further consideration. 
Our LLM presentation nudge meets a variety of theoretical nudging definitions: a smart nudge to guide cognitive process~\cite{mele_smart_2021}, a nudge to reflect on the consequences of past choices~\cite{sunstein_nudging_2014}, and a self-actualization nudge to help users develop new tastes~\cite{knijnenburg2016recommender, mattis_nudging_2024}.
Consistent with prior work~\cite{mattis_nudges_2025, mattis_nudging_2025}, our aggregated results show that manipulating the information presented to the DC-NP group was largely ineffective.
Considering how the nudging framework distinguishes between algorithmic and presentation nudges, we note the intractability of disentangling these interventions in practice. 
For example, recommendation position/rank had the strongest impact on engagement, but this nudge operates simultaneously on exposure and selectivity, which conflates whether diverse consumption resulted from genuine selectivity or exposure constraints.
As shown in Figure~\ref{fig:prompt_engagement}, event-based rewrites led to large, but marginally significant, increases in click likelihood, with a trend towards reducing users' prior topic interest (i.e., their habitual consumption patterns).
Future work should integrate presentation nudging goals into ranking algorithms to increase the number of legitimate event-based narrative rewrites and the statistical power of this intervention.
This approach offers a promising direction for impacting the selectivity component of diversity goals, enabling users to exercise agency in actualizing diverse decisions.

\subsection{Social Impact and News Integrity}




News recommendation calibration can continuously select articles of different topics and locality for readers to consume, shifting their information perception and understanding of things happening \cite{flaxman2016filter,helberger2021democratic}. Our survey responses reveal a normative shift: users in the treatment groups maintained or increased their belief that newsletters should provide variety, whereas control group agreement dropped. This indicates that long-term exposure to calibrated diversity may cultivate a more civically-minded reader who expects and values a diverse news perspective, consistent with the virtuous circle hypothesis \cite{stromback2010media,helberger2021democratic}.  
While personalized previews (DC-NP) did not significantly alter consumption diversity compared to calibration alone, they played a vital role in user agency. We found that participants in the DC-NP group reported a higher perception of control over their recommendations. This suggests that the social impact of LLM-driven personalization may be less about changing what people read and more about increasing the transparency and legibility of the algorithm \cite{tintarev2015explaining,diakopoulos2025prospective}. By explaining the "why" behind a recommendation through personalized context, we move from black-box nudging towards a collaborative discovery process between users and models \cite{heer2019agency,shneiderman2020human}. Nonetheless, one key challenge we observed while experimenting with news preview rewrite was the balance between original news content factuality and personalized context integration. Specifically, the selection of the similarity threshold between the context article and the target news article required multiple pilot rounds to avoid over-personalization that could distort the original reporting.

Future work should investigate more robust safeguard mechanisms to ensure that the personalized hook does not come at the cost of news integrity, such as the "Chain-of-verification" \cite{dhuliawala2024chain} and "LLM-as-a-judge" \cite{gu2024survey}. While our current approach relied on heuristic similarity thresholds, future iterations could employ secondary fact-check models to verify that the core event details and news entity remain uncompromised during the rewrite process. Additionally, exploring the trade-offs between event-based and topic-based rewrites could help determine the optimal granularity of personalization that maximizes engagement without straining the computational resources or the factual boundaries of the news ecosystem \cite{bubeck2023sparks}. Future work should further refine these presentation nudges and their factual safeguards, ultimately supporting algorithms to make more novel connections between diverse news articles.

\section{Limitations}\label{sec:limitations}
We acknowledge three major limitations for this study. First, the taxonomy of news topic and locality is defined by AP editors and their automated tagging algorithms. In particular, a small amount of news is categorized as \textit{"Neither"} for its locality and can occasionally limit the balanced locality calibration. Future works can further explore more granular news geography calibration, such as local, regional, national, continental, and world level news. 

Second, as POPROX is a relatively new platform, our study was impacted by the typical lifecycle of an emerging tool. This manifested in two primary ways: a relatively small sample size of highly active users and a trend of gradual churn or lower engagement over time as the initial novelty for newly subscribed users diminished. These factors, plus the stationarity of response variables and windowing of experiment data, may limit the generalizability of our longitudinal findings to larger, more established news environments.\looseness=-1

Finally, the ecological setting of POPROX presented challenges for data collection. As an ethical research platform, POPROX does not collect behavioral logs beyond API calls (i.e., email opens and reading behavior), limiting our ability to track more granular engagement behaviors. Moreover, because the platform relies entirely on participants' intrinsic motivation, their participation in our weekly and study-specific surveys was voluntary. This resulted in a limited volume of survey responses, which restricted our ability to achieve between-subject statistical significance for certain perceptual metrics. While these survey results provided valuable observational trends to complement our log analysis, future studies might benefit from a hybrid incentive model to ensure more consistent longitudinal feedback without compromising the naturalistic setting of the news consumption experience.


\section{Conclusion}
In this work, we clarify the relationship between exposure and consumption diversity by empirically evaluating two types of nudges on domestic and world news coverage.
Through a 5-week longitudinal study with 120 U.S. readers on the POPROX platform, we demonstrate that a topic-locality dual-calibration algorithmic nudge successfully expands both exposure and consumption diversity across domestic and world news coverage.
For LLM-based presentation nudges, our findings reveal important nuances in the impact of personalization strategies: a highly specific, event-based approach outperforms both topic-based and no personalization approaches, but the lack of opportunities for legitimate event-based rewrites limits its impact on consumption diversity.
These results highlight event-based contextualization as a promising direction for future work to influence individual selectivity when faced with diverse recommendations.
Together, this work contributes to the limited body of research examining the longer-term effects of diversity nudges on users' situated reading behavior.

\bibliographystyle{ACM-Reference-Format}
\bibliography{main}

\clearpage
\onecolumn
\appendix
\section{Appendix}

\subsection{Prompt Details}
\label{app:prompt}

\begin{figure}[H]
    \centering
    \includegraphics[width=1.0\linewidth]{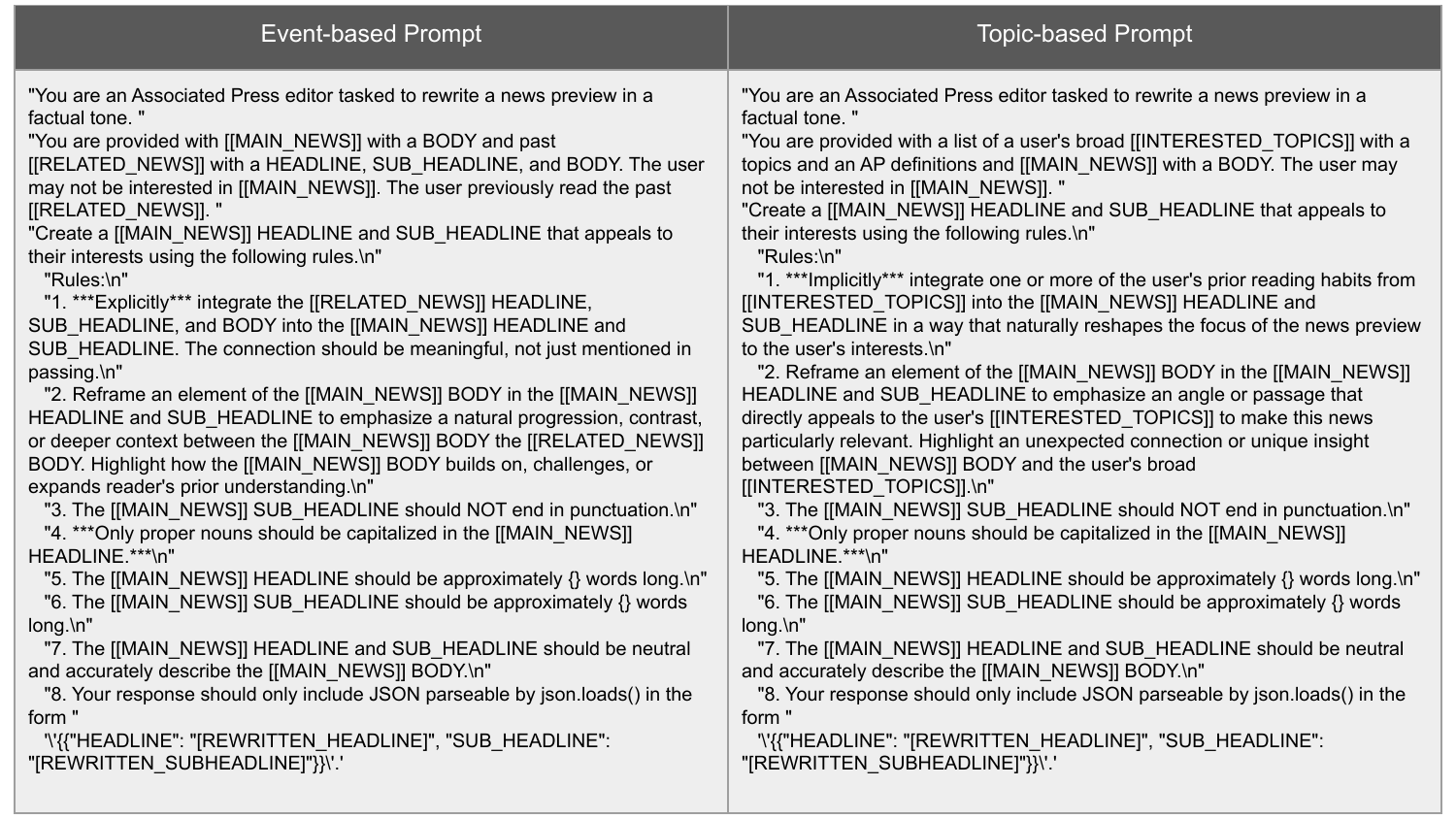}
    \caption{Event- and topic-based news rewrite prompt.}
    \label{fig:detail-prompt}
\end{figure}

\clearpage

\subsection{Pre- and Post-Survey Questions} \label{appendix:survey-questions}

\begin{table}[H]
\centering
\small 
\renewcommand{\arraystretch}{1.3} 
\begin{tabularx}{\textwidth}{l X p{4.5cm} l}
\toprule
\textbf{ID} & \textbf{Question Text} & \textbf{Response Options} & \textbf{Phase} \\ 
\midrule
Q1 & I feel that it is important that a newsletter provides a variety of international and domestic news topics to its readers. & 5-point Likert (Strongly Disagree [SD] -- Strongly Agree [SA]) & Pre \& Post \\ 

Q2 & I feel that it is important that people read a variety of international and domestic news topics. & 5-point Likert (SD -- SA) & Pre \& Post \\ 

Q3 & I think that personalizing the text of news headlines and descriptions to my interests improves my news reading experience. & 5-point Likert (SD -- SA) & Pre \& Post \\ 

Q4 & I think that using artificial intelligence (AI) to personalize news previews to my interests is accurate and trustworthy. & 5-point Likert (SD -- SA) & Pre \& Post \\ 

Q5 & In recent newsletters you received, reflect on the relevance of the articles to your interests. About how many articles were relevant to your interests? & None, 1--3, 4--6, 7--9, All of them & Pre \& Post \\ 

Q6 & In recent newsletters you received, reflect on the balance of international and domestic articles. I noticed\dots & A lot more intl., A little more intl., No difference, A little more domestic, A lot more domestic, I’m not sure & Pre \& Post \\ 
\midrule
Q7 & \textit{[If Q6 $\neq$ ``Not sure'']} This balance was intended to recommend a similar ratio of international and domestic news based on our source provider. Reflecting on the balance, do you think that's an improvement? & 5-point Likert (SD -- SA) & Post Only \\ 

Q8 & I would like the future POPROX newsletters to also balance international and domestic news. & 5-point Likert (SD -- SA) & Post Only \\ 
\midrule
Q9 & I feel like I have control over what articles POPROX News selects. & 5-point Likert (SD -- SA) & POPROX native, Pre \& Post\\
\bottomrule
\end{tabularx}
\caption{Survey Questions and Response Scales}
\label{tab:survey_questions}
\end{table}

\end{document}